\documentstyle[12pt,preprint,aps,pre,epsf]{revtex}

\newcommand{\myfig}[3] {
\begin{figure}
\centerline{
\epsfysize=#2  \epsfbox{#1.eps}
}
\caption{#3}
\label{#1}
\end{figure} }

\def\be{ \begin{equation} }
\def\ee{ \end{equation} }

\begin{document}
\tightenlines
\draft
\title{\bf
Finite Size Scaling of the 2D Six-Clock model}
\author{M.~Itakura}

\address{Center for Promotion of Computational Science and Engineering,
Japan Atomic Energy Research Institute,
Meguro-ku, Nakameguro 2-2-54, Tokyo 153, Japan\\}
\date{\today}
\maketitle

\begin{abstract}
We investigate 
the isotropic-anisotropic phase transition of
the two-dimensional XY model with six-fold anisotropy,
using Monte Carlo renormalization group method.
The result indicates difficulty of
observing asymptotic critical behavior
in Monte Carlo simulations, 
owing to the marginal flow at the fixed point.

\end{abstract}

The XY model with $Z_6$ symmetry breaking field
represents not only planar spin magnets with $Z_6$
symmetric crystal field \cite{z6},
but also models with $Z_2 \times Z_3$ symmetry
(which is isomorphic to $Z_6$) 
such as 
three-state antiferromagnetic Potts model on the square and the cubic
lattice  \cite{potts} and
Ising antiferromagnet on the triangular and
the stacked-triangular lattice 
\cite{ising3d,landau,kitatani,miyashita,queiroz,fss}.

The critical behavior of the model in two dimension is well understood
by scaling argument \cite{z6}:
The model undergoes two distinct phase transitions, both of
which being Kosterlitz-Thouless transition.
In Monte Carlo simulations of finite system, however,
marginal renormalization flow near the fixed point
makes it difficult to observe asymptotic critical behavior.
In the present work,
we used improved Monte Carlo renormalization group method \cite{mcrg}
to observe the renormalization flow
of the six-clock model
on the square lattice of size up to $64\times 64$,
and found that the size $64\times 64$ is still insufficient to
observe asymptotic critical behavior.
Similar situation occurs in several other models, such as
general spin models in $D=4$ \cite{mcrg}
and $O(3)$ spin model with cubic anisotropy in $D=3$ \cite{cubic}.

In the Ref.\cite{z6} by Jos\'{e} $et.al.$, 
they investigated the  following model:
\be
H=K \sum_{<ij>}\cos(\theta_i-\theta_j) + A \sum_i \cos(p \theta_i)
\label{H}
\ee
where the first summation runs over all nearest pairs on the square lattice, 
and $p$ is some integer.
From Gaussian spin-wave theory and scaling analysis,
it can be shown that the perturbation term $\sum_i \cos(p \theta_i)$
is relevant when $\eta<4/p^2$, where $\eta$ is the critical exponent
of long distance spin-spin correlation defined as below:

\be
<\cos(\theta_x - \theta_y)>\sim |x-y|^{-\eta} \,\,\,
{\mathrm for}\,\,\,\,\,\,\, |x-y| \gg 1.
\ee

Thus flow of the renormalized parameters
$K$ and $A$ for
$p=6$ case is expected to
become the one shown in Fig. \ref{rgkt}: 
The parameter space is divided into three regions,
namely high-temperature (H), Kosterlitz-Thouless (KT), and
low-temperature (L) phases.
Any model whose initial parameter crosses the L-KT boundary
is attracted to a fixed point which is on the KT fixed line 
and characterized by an exponent $\eta=1/9$ (we denote the fixed point
 as $F_{1/9}$)
and exhibits KT-phase transition.
Near the fixed point $F_{1/9}$,
renormalization flow along the L-KT boundary line
slows down 
because $F_{1/9}$ is marginal for both direction
$K$ and $A$.
Thus finite size correction (distance to the final fixed point)
is expected to behave like $1/\log L$ ($L$ is linear size of
the system)  
and one can not observe
asymptotic critical behavior unless extremely large system
is used.
%%%%%%%%%%%%%%%%%%hoge
For example, the critical exponent $\eta$ at the lower critical temperature
of Ising antiferromagnet on the triangular lattice
has been estimated by several authors
(to confirm theoretical prediction $\nu=1/9=1.111\cdots$)
as $\eta=0.15(2)$ \cite{landau}
and $\eta=0.125(25)$ \cite{miyashita}. 
The accuracy of these values are relatively low compared to other
models such as $O(n)$ spin models in three dimension:
this implies that there are large finite-size correction.

In the present work we numerically investigated the renormalization
flow of the model (\ref{H})
using the improved Monte Carlo Renormalization Group (MCRG)
method \cite{mcrg}, which is very simple and efficient way to
extract essential information of critical phenomena from
simulation data.
We observed
the following quantities:
\be
K_L=1-{ <{\bf S}({\bf k}_1)\cdot {\bf S}(-{\bf k}_1)>\over<{\bf S}(0)^2>}
, \,\,\,\,
A_L= {<R_M^6 \cos(6\theta_M)>\over <R_M^2>^3}
\ee
where
\be
{\bf S}({\bf k}) \equiv L^{-2}\sum_{\bf x} \exp (i {\bf k}\cdot {\bf x})
(\cos\theta_{\bf x},\,\, \sin \theta_{\bf x} ),
\,\,\,\,\,
{\bf k}_1 \equiv (2\pi/L,0),
\ee
and
\be
{\bf M}\equiv {\bf S}(0)=R_M (\cos \theta_M ,\,\,\,\sin \theta_M).
\ee
$K_L$ and $A_L$ 
reflect the behavior of the renormalized 
temperature and anisotropy (by a factor $L$), respectively.
We also observed Binder's parameter and $<\cos 6 \theta_M>$,
and found that $K_L$ and $A_L$ reflect the RG flow better
than these quantities.

The Hamiltonian (\ref{H}) on a 
$L\times L$ square lattice is simulated for $L=16,32,$ and $64$.
In Monte Carlo simulation, Metropolis update scheme
and Wolff's cluster algorithm \cite{wolff}
are combined.
In the Metropolis update,
we choose new spin value $\theta$ 
with a probability proportional to
$\exp [-A \cos(6 \theta)]$ and calculate
acceptance probability using only $K \sum_{<ij>}\cos(\theta_i-\theta_j)$.
This scheme satisfies the detailed-balance condition of the 
Hamiltonian (\ref{H}) and improves the acceptance ratio for large $A$.
In the cluster update, spin-reflection axis is restricted to 
the one which preserves the anisotropy.
The Hamiltonian
(\ref{H}) is simulated for $A=\infty, 0.5, 0.2, 0.1,$ and $0.05$ cases.

Figure \ref{rgmc} shows the flow of $K_L$ and $A_L$: each line is drawn from
$(K_L, A_L)$ to $(K_{2L},A_{2L})$.
One can see that the obtained RG flow agrees well with the theoretical
one.
The position of the critical point $F_{1/9}$ is estimated by
plotting $\log <M_L^2> + (\log L)/9$ versus $\log L$ for
various $K$, being $A$ fixed to zero:
since $<M_L^2> \sim L^{-\eta}$, the plot becomes horizontal
at $F_{1/9}$.
Fig. \ref{g9} indicates that $F_{1/9}$ is located
near $K=1.7$. The value of $K_L$ at this temperature
is marked as $F_{1/9}$ in Fig. \ref{rgmc}.
One can see that the RG flow is attracted to the KT line 
in the left (high-temperature) side of  $F_{1/9}$,
while the flow deviates from the KT line in the right (low-temperature)
side of $F_{1/9}$.
Plots for $A=\infty$ case seems to approach $F_{1/9}$ as
$L$ increases. However, the approach is indeed slow 
and extremely large $L$ is required to observe convergence
to $F_{1/9}$.

The distance to $F_{1/9}$ results in systematic error in
the finite-size scaling analysis 
\cite{landau,miyashita}, such as position of the
critical point and value of the critical exponent,
and one can not extrapolate the $L=\infty$ limit
owing to the slow vanishing correction term.
For example, if one define finite-size critical point
as a temperature where 
renormalized anisotropy becomes size-independent,
it will deviate to low-temperature side as Fig. \ref{rgkt}
suggests.
Similarly, the $\eta=1/9$ criterion will lead to
systematic error, because effective exponent for $\eta$
at $A\neq 0$ region generally differ from that of 
asymptotic value at $A=0$.
%However, observation of renormalized anisotropy gives clearer
%qualitative information, compared to the $\eta=1/9$ criterion.

In conclusion, we demonstrated that
it is nearly impossible, by means of Monte Carlo simulations
of finite system to observe
asymptotic critical behavior
of isotropic-anisotropic phase transition in 2D six-clock model
and those of similar symmetry.
Although {\it quantitative} information
such as position of critical point or values of critical 
exponents are hard to obtain,
{\it qualitative} information such as
presence or absence of
the transition can be easily obtained
by observing renormalized anisotropy, and
it is enough for Monte Carlo simulation since the nature of the
transition, such as critical exponents, is already well known.

\myfig{rgkt}{8cm}{Theoretical prediction of the renormalization flow
of temperature and anisotropy. L, KT, and H denotes
low-temperature, Kosterlitz-Thouless, and high-temperature phase,
respectively. Dotted lines indicate direction of
marginal renormalization flow.}

\myfig{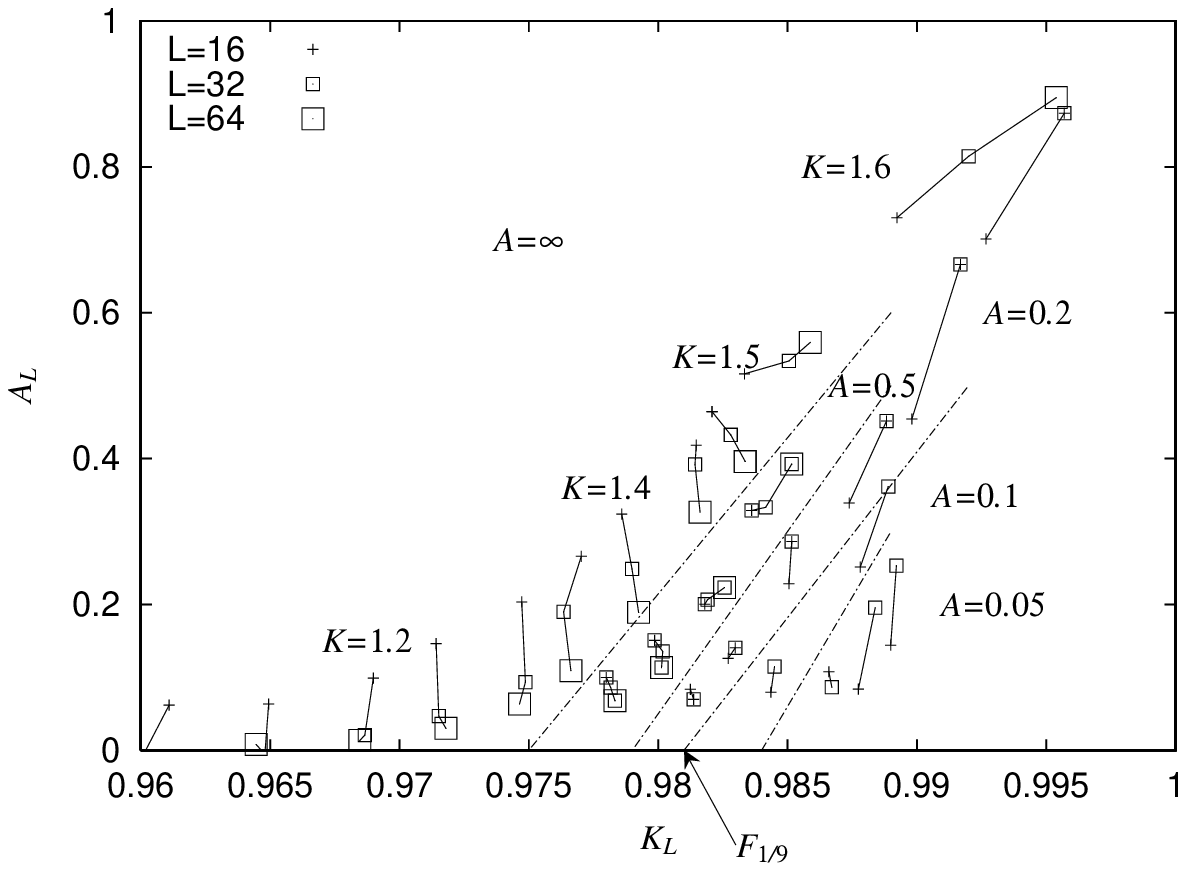}{8cm}{Renormalization flow obtained from Monte Carlo
simulations. 
Connected two line segments are drawn from $L=16$ data to $L=32$ data,
then to $L=64$ data.  Other single lines are drawn from 
$L=16$ data to $L=32$ data.
Plots for each different values of $A$ are separated by
dotted lines.
The position of $F_{1/9}$ was estimated from finite-size scaling of
$<M_L^2>$.
}

\myfig{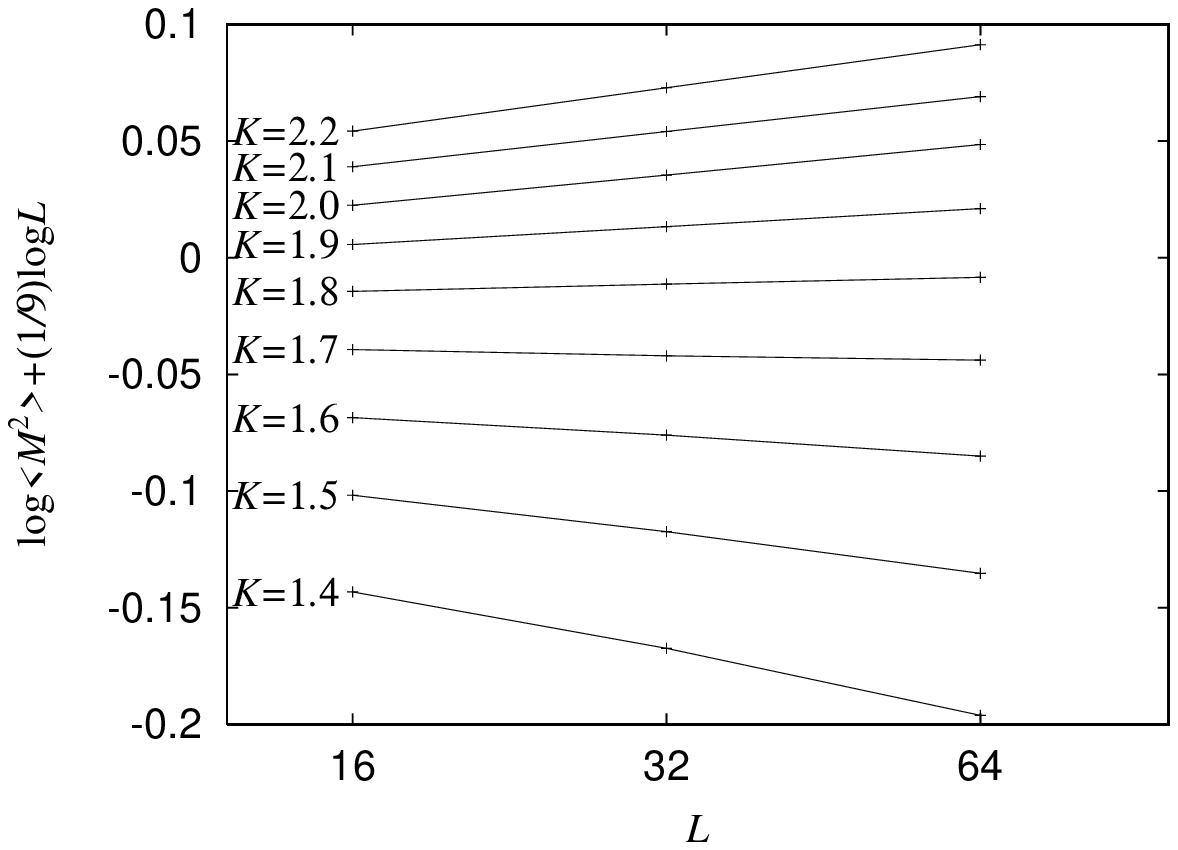}{8cm}{Plot of $\log <M_L^2>+{1\over 9}\log L$ versus $\log L$
for $A=0$.}

\end{document}